\documentclass[english]{article}
\usepackage[T1]{fontenc}
\usepackage[latin9]{inputenc}
\usepackage{textcomp}
\usepackage{amsmath}
\usepackage{amssymb}

\makeatletter

\DeclareRobustCommand{\greektext}{%
  \fontencoding{LGR}\selectfont\def\encodingdefault{LGR}}
\DeclareRobustCommand{\textgreek}[1]{\leavevmode{\greektext #1}}
\DeclareFontEncoding{LGR}{}{}
\DeclareTextSymbol{\~}{LGR}{126}
\newcommand{\lyxmathsym}[1]{\ifmmode\begingroup\def\b@ld{bold}
  \text{\ifx\math@version\b@ld\bfseries\fi#1}\endgroup\else#1\fi}

\newcommand{\lyxaddress}[1]{
\par {\raggedright #1
\vspace{1.4em}
\noindent\par}
}

\makeatother

\usepackage{babel}
\begin{document}

\title{Energy levels and extension of the Schrödinger operator.}

\author{Y. C. Cantelaube%
\thanks{e-mail : yves.cantelaube@univ-paris-diderot.fr%
}}

\maketitle

\lyxaddress{U.F.R. de Physique, Université Paris Diderot, Bâtiment Condorcet,
75205 Paris cedex 13, France}
\begin{abstract}
Although energy levels are often given by solutions of the radial
equation such that u(0) is non zero, and hence by first-order singular
functions which are not eigenfunctions of H, the latter is always
considered as the only operator that gives energy levels. Vibrational
levels of diatomic molecules are a usual example. We show that the
operator which has singular eigenfunctions, or pseudofunctions, that
give energy levels, is the operator whose action on pseudofunctions
amounts to the embedding in the distributions of R$^{3}$ of their
Hamiltonian in R$^{3}$/\{0\}. When its eigenfunctions are regular,
this operator amounts to H. Energy levels, which are given by eigenfunctions
of H when u(0) is zero, are thus given in any case by eigenfunctions
of this operator, which is an extension of the Schrödinger operator,
but not of the Hamiltonian.
\end{abstract}

\section{Introduction. Energy levels and radial equation.}

In a central potential $\mathcal{V}$(\textit{r}) energy levels $\mathcal{E}{}_{n}$
are eigenvalues of the Hamiltonian that belong to normalized eigenfunctions
$\psi_{n}$ = {[}$\mathit{w_{n}}\left(r\right)$/\textit{r}{]}$Y_{\ell}^{\mu}\left(\theta,\varphi\right)$
which behave at the origin like\textit{ r}$^{\ell}$$Y_{\ell}^{\mu}\left(\theta,\varphi\right)$,
and hence eigenvalues of the radial equation that belong to normalized
eigensolutions $\mathit{w_{n}}\left(r\right)$ which behave at the
origin like\textit{ r}$^{\ell+1}$,

\begin{equation}
\left[-\frac{\hbar^{2}}{2m}\frac{d^{2}}{dr^{2}}+\frac{\ell(\ell+1)\hbar^{2}}{2mr^{2}}+\mathcal{V}(r)\right]w_{n}\left(r\right)=\mathcal{E}{}_{n}\mathit{w_{n}\left(r\right)}
\end{equation}

As physical potentials $\mathcal{V}$(\textit{r}) are not exactly
known, one must substitute theoretical and then approximate potentials
\textit{V}(\textit{r}) for which the Hamiltonian has a discrete set
of normalizable eigenfunctions $\Psi_{n}$= {[}$\mathit{u_{n}}\left(r\right)$/\textit{r}{]}$Y_{\ell}^{\mu}\left(\theta,\varphi\right)$,
and hence the radial equation a discrete set of normalizable analytic
eigensolutions $\mathit{u_{n}}\left(r\right)$which behave at the
origin like\textit{ r}$^{\ell+1}$,

\begin{equation}
\left[-\frac{\hbar^{2}}{2m}\frac{d^{2}}{dr^{2}}+\frac{\ell(\ell+1)\hbar^{2}}{2mr^{2}}+V\left(r\right)\right]u_{n}\left(r\right)=E_{n}\mathit{u_{n}\left(r\right)}
\end{equation}

One assumes indeed that if \textit{V}(\textit{r}) $\mathcal{\approx V}$(\textit{r}),
$\Psi_{n}$ will be similar to $\psi_{n}$, $\mathit{u_{n}}\left(r\right)$
to $\mathit{w_{n}}\left(r\right)$, and $\mathit{E_{n}\approx\mathcal{E}{}_{n}}$.
The solutions of the radial equations, which are written in the form
of series, $\mathit{u_{n}\left(r\right)}$ = $\mathit{r^{\lambda}}$$\sum_{k\geq0}{\textstyle a_{k}^{n}r^{k}}$,
are given in $\textrm{R}^{3}$/\{0\} by both roots $\lambda$= $\ell+1$,
and $\lambda$= $-\ell$, but in R$^{3}$ for $\ell$ > 0 only by
the root $\lambda$= $\ell+1$ {[}1{]}. These solutions are normalizable
and give, when substituted in $\Psi_{n}$, eigenfunctions of \textit{H},
or solutions of the Schrödinger equation. For $\ell$ = 0 the radial
equations
\begin{equation}
\left[-\frac{\hbar^{2}}{2m}\frac{d^{2}}{dr^{2}}+\mathcal{V}\left(r\right)\right]w_{n}\left(r\right)=\mathcal{E}{}_{n}\mathit{w_{n}\left(r\right)}
\end{equation}

\begin{equation}
\left[-\frac{\hbar^{2}}{2m}\frac{d^{2}}{dr^{2}}+V\left(r\right)\right]u_{n}\left(r\right)=E_{n}\mathit{u_{n}\left(r\right)}
\end{equation}
have solutions in $\textrm{R}{}^{3}$ given by both roots, that is,
$\lambda$ = 1, and $\lambda$= 0. Both roots give normalizable solutions,
but only the solutions given by the root $\lambda$ = 1 give, when
substituted in $\Psi_{n}$, solutions of the Schrödinger equation
{[}1{]}. Insofar as energy levels $\mathcal{E}{}_{n}$ are eigenvalues
of the Hamiltonian, the solutions $\mathit{w_{n}}\left(r\right)$
of (3) behave at the origin like \textit{r}. In order for $\Psi_{n}$
to be eigenfunctions of \textit{H}, one must substitute potentials
\textit{V}(\textit{r}) for which the following conditions must be
satisfied: \textit{i}) \textit{V}(\textit{r}) must be close to $\mathcal{V}\left(r\right)$,
at least in the neighborhood of the minimum; \textit{ii}) Eq.(4) must
have a discrete set $\mathit{u_{n}}\left(r\right)$ of normalizable
analytic solutions ; \textit{iii}) these solutions must behave at
the origin like \textit{r}.

Now, explicit analytic solutions of (4) can be derived only for a
limited number of potential energy functions \textit{V}(\textit{r}).
When we demand that the first two conditions be satisfied, this number
is much more limited. If we moreover demand that these solutions behave
at the origin like \textit{r}, the problem cannot necessarily be solved.
This is why one substitutes potentials \textit{V}(\textit{r}) close
to $\mathcal{V}\left(r\right)$, for which (4) has a discrete set
of normalizable solutions \textit{regardless of their value at the
origin}. One intuitively expects indeed that if \textit{$V\left(r\right)\approx\mathcal{V}\left(r\right)$}
and if (4) has a discrete set of normalizable solutions $\mathit{u_{n}}\left(r\right)$,
moreover demanding that they behave at the origin like \textit{r}
is a \textit{supplementary} condition that is \textit{not required}
to obtain eigenvalues $\mathit{E_{n}}$ of (4) close to the eigenvalues
$\mathcal{E}{}_{n}$ of (3). This is shown by the following example.

\section{Vibrational levels of diatomic molecules.}

The vibrational levels of diatomic molecules are of the form

\[
\mathcal{E}{}_{n}=-\mathit{V_{m}+(n+\text{\textonehalf)}}\hbar\omega+C_{2}(n+\text{\textonehalf)}^{2}+C_{3}(n+\text{\textonehalf)}^{3}+...\qquad n=0,1,2...
\]

In the Born-Oppenheimer approximation the potential energy of interaction
between the two nuclei of a diatomic molecule is a central potential
\textit{V}(\textit{r}), attractive at large distances, repulsive at
short distances, with a minimum at \textit{r} = $\mathit{r_{m}}$.
By expanding the potential in powers of \textit{r} \textendash{} $\mathit{r{}_{m}}$
in the neighborhood of the minimum by neglecting the terms of order
$\geq$ 3, it gives

\begin{equation}
V\left(r\right)=\text{\textonehalf}m\omega{}^{2}\left(r-r_{m}\right)^{2}-V_{m}\qquad\qquad\omega=[V\,"(r_{m})/m]^{1/2}
\end{equation}

This parabolic potential is used as a first approximation, which holds
in the neighborhood of the minimum, and then gives the first levels.
The radial equation for $\ell=0$ involving this potential,

\begin{equation}
\left[-\frac{\hbar^{2}}{2m}\frac{d^{2}}{dr^{2}}+\text{\textonehalf}m\omega{}^{2}\left(r-r_{m}\right)^{2}-V_{m}\right]u_{n}\left(r\right)=E_{n}\mathit{u_{n}\left(r\right)}
\end{equation}

has the square integrable solutions and the corresponding eigenvalues

\[
u_{n}\left(r\right)=N_{n}exp\left[-\text{\textonehalf}\beta^{2}(r-r_{m})^{2}\right]H_{n}\left[\beta(r-r_{m})\right]\qquad\beta^{2}=m\omega/\hbar
\]

\begin{equation}
E_{n}=(n+\text{\textonehalf)}\hbar\omega-V_{m}\qquad\qquad\qquad n=0,1,2...
\end{equation}

where $\mathit{H_{n}}$ are Hermite polynomials of order \textit{n}
and $\mathit{N_{n}}$ normalization constants. Experimental data show
that these eigenvalues give estimates of the first vibrational levels
of most molecules. Now, except for the case where $\beta r_{m}$ coincides
with a zero of $\mathit{H_{n}}$,

\begin{equation}
u_{n}\left(0\right)=N_{n}exp(-\text{\textonehalf}\beta^{2}r_{m}{}^{2})H_{n}(\beta r_{m})\neq0
\end{equation}

Analytical forms for \textit{V}(\textit{r}) closer to the realistic
physical potential have been proposed, the most frequently used is
the Morse potential {[}2{]}

\begin{equation}
V_{M}(r)=V_{m}\{exp\left[-2a(r-r_{m})\right]-2exp\left[-a(r-r_{m})\right]\}
\end{equation}

For \textit{$\ell$} = 0 the square-integrable solutions and the eigenvalues
of the radial equation are {[}2{]}

\[
u_{n}(r)=exp\left[-z(r)/2\right]\left[z(r)\right]^{b/2}L_{n+b}^{b}\left[z(r)\right]
\]

\[
z=2d\, exp\left[-a(r-r_{m})\right]\qquad d=(2mV_{m})^{\lyxmathsym{\textonehalf}}(a\hbar)^{-1}\qquad b=2d-1-2n
\]

\[
E_{n}=(n+\text{\textonehalf)}\hbar-\left(n+\text{\textonehalf}\right)^{2}(\hbar^{2}\omega^{2}/4V_{m})-V_{m}\qquad\omega=a(2V_{m}/m)^{\text{\textonehalf}}
\]
where $\mathit{L_{n+b}^{b}}\left(z\right)$ are generalized Laguerre
polynomials. Experimental data show that these eigenvalues give very
accurate values for the vibrational levels of nearly all molecules,
but again the condition $u{}_{n}$(0) = 0 is not satisfied.

If one considers that by substituting the potential (5) \textquotedblleft{}we
are left with the equation of a one-dimensional harmonic oscillator
centered at\textit{ r} = $\mathit{r_{m}}$'' {[}3,$\textrm{A}_{\textrm{V}}${]},
no boundary condition is required at the origin, so that in particular
perturbation theory for a one-dimensional harmonic oscillator is applied
to the third-order term.

Otherwise, arguments have been proposed to justify that the fact that
$u{}_{n}(0)\neq0$ is of no significance. According to Pauling \textit{et
al}. the solutions of the radial equation must be zero at \textit{r}
= 0 and +$\infty$, whereas the solutions of the harmonic oscillator
must be zero at \textit{r} = $\text{\textendash}\infty$ and +$\infty$,
but because of the rapid decrease in the harmonic oscillator functions
outside the classically permitted region, ``it does not introduce
a serious error to consider that the two sets of boundary conditions
are practically equivalent\textquotedblright{} {[}4, See also 5,6,7{]}.
As noted by Schiff, the eigenvalues of the radial equation are those
of a linear harmonic oscillator \textquotedblleft{}if the domain of
\textit{r} is extended to $-\infty$\textquotedblright{} {[}8{]},
which amounts to substituting a one-dimensional problem for a three-dimensional
problem. Or else, one considers that, as (6) is the radial equation
for zero angular momentum states, the exact solutions must be rigorously
zero at \textit{r} = 0, but the solutions (7) are ``practically zero
at the origin'' {[}3,$\textrm{F}{}_{\textrm{VII}}${]}. It should
be noted that one often considers indeed that the exact solutions
of (6) can be obtained only if the variable is taken from $-\infty$
to +$\infty$, or if the solutions vanish at the origin. In fact,
(7) are the exact solutions of (6), that is, the solutions in R$^{3}$
($r\geq$0), as is easily checked.

Similarly, the value of the Morse potential is sometimes considered
as so large at \textit{r} = 0 that the Morse eigenfunctions are ``effectively
zero for \textit{r} < 0\textquotedblright{} {[}9{]}. In fact they
are non zero at \textit{r} = 0, but according to Morse since \textquotedblleft{}in
every case \textit{r}\textgreek{Y} will be extremely small \ldots{}
this discrepancy will not affect the values of the energy levels\textquotedblright{}
{[}2, See also 6,7{]}.

But if the two boundary conditions, $u_{n}\left(0\right)$ = 0 and
$u{}_{n}\left(0\right)\approx0$, are practically equivalent, it merely
means that the former condition is not required. First-order singular
functions are not indeed ``practically eigenfunctions'' of \textit{H},
they are eigenfunctions of \textit{H} in $\textrm{R}^{3}$/\{0\},
but not in $\textrm{R}^{3}$. In other words, whether or not \textit{u}(0)
is very small, if it is non zero, \textgreek{Y}$_{n}$ behaves at
the origin like 1/\textit{r}, and what these examples show is thus
that energy levels are not affected because that they are not given
by eigenfunctions of \textit{H}. By \textit{changing the condition
at the origin, we change indeed the operator}.

\section{The energy levels and the operator $\mathit{H_{d}}$.}

We must thus determine the operator which has singular eigenfunctions
of the form $\Psi_{n}$ = {[}$\mathit{u_{n}}\left(r\right)$/\textit{r}{]}$Y_{\ell}^{\mu}\left(\theta,\varphi\right)$
and corresponding eigenvalues $\mathit{E_{n}}$, such that for $\ell$
= 0 $\mathit{u_{n}\left(r\right)}$ and $\mathit{E_{n}}$ are solutions
of (4) with $u_{n}\left(0\right)\neq0$. Now, when $\Psi_{n}$ and
$u_{n}$(\textit{r}), which are of $\textrm{C}^{\infty}(\textrm{R}$$^{3}$/\{0\}),
are singular at the origin, they define in R$^{3}$ distributions
called pseudofunctions, denoted by $\textrm{Pf}.\Psi_{n}$ and $\textrm{Pf}.u_{n}$(\textit{r})
{[}10{]}, and the latter are not a solution of (2) {[}1{]}. The radial
equation which has in R$^{3}$ singular solutions $\textrm{Pf}.u_{n}$(\textit{r})
given by the root $\lambda$ = $-\ell$ is the \textit{extension}
in R$^{3}$ of (2), it is the equation {[}1{]}
\begin{equation}
\textrm{Pf}.\left[-\frac{\hbar^{2}}{2m}\frac{d^{2}}{dr^{2}}+\frac{\ell(\ell+1)\hbar^{2}}{2mr^{2}}+V\left(r\right)\right]u_{n}\left(r\right)=E_{n}\textrm{Pf}.\mathit{u_{n}\left(r\right)}
\end{equation}

Its solutions, given by both roots $\lambda$ = $\ell+1$ and $\lambda$
= $-\ell$, are either functions $\mathit{u_{n}}\left(r\right)$ which
behave at the origin like $r{}^{\ell+1}$, or pseudofunctions $\textrm{Pf.}\mathit{u_{n}}\left(r\right)$
which behave at the origin like $\textrm{Pf.}r^{-\ell}$. (The symbol
Pf. is used when we have either the former or the latter). When the
solutions of (10) are given by the roots $\lambda$ = $\ell+1$ and
$\lambda$ = 0 (i.e. \textit{$\lambda$} = $-\mathit{\ell}$ for $\ell$
= 0), the symbol Pf. is useless and (10) is written in the form (2).
Just as $\Psi_{n}$ defines in R$^{3}$ the pseudofunction $\textrm{Pf}.\Psi_{n}=\textrm{Pf.}[\mathit{u_{n}}\left(r\right)/r]Y_{\ell}^{\mu}\left(\theta,\varphi\right)$,
its Hamiltonien in R$^{3}$/\{0\}, which is the Hamiltonian in the
sense of the functions

\[
H\Psi_{n}=\frac{1}{r}\left[-\frac{\hbar^{2}}{2m}\frac{d^{2}}{dr^{2}}+\frac{\ell(\ell+1)\hbar^{2}}{2mr^{2}}+V(r)\right]\mathit{u_{n}}(r)Y_{\ell}^{\mu}\left(\theta,\varphi\right)\qquad\qquad r>0
\]
defines in R$^{3}$ the distribution

\begin{equation}
\textrm{Pf}.H\Psi_{n}=\textrm{Pf}.\left\{ \frac{1}{r}\left[-\frac{\hbar^{2}}{2m}\frac{d^{2}}{dr^{2}}+\frac{\ell(\ell+1)\hbar^{2}}{2mr^{2}}+V(r)\right]\mathit{u_{n}}(r)Y_{\ell}^{\mu}\left(\theta,\varphi\right)\right\} 
\end{equation}

Let $\mathit{H_{d}}$ the operator whose action on a pseudofunction
Pf.$\Psi_{n}$ amounts to the embedding in the distributions of R$^{3}$
of the Hamiltonian of $\Psi_{n}$ in R$^{3}$/\{0\},

\begin{equation}
\mathit{H_{d}}\textrm{Pf}.\Psi_{n}=\textrm{Pf}.H\Psi_{n}
\end{equation}

If Pf.$\mathit{u_{n}\left(r\right)}$ is a solution of (10), whether
it is given by the root \textit{$\lambda$} = $\mathit{\ell}$ + 1,
or by the root \textit{$\lambda$} = $-\mathit{\ell}$, by substituting
(10) in (12), we obtain the eigenvalue equation of $\mathit{H_{d}}$,
\begin{equation}
H_{d}\textrm{Pf.}\Psi_{n}=E_{n}\textrm{Pf.}\Psi_{n}
\end{equation}

where $\mathit{E_{n}}$ is the eigenvalue of (10) that belongs to
Pf.$\mathit{u_{n}}\left(r\right)$.

The \textit{normalizable} solutions are given by the roots \textit{$\lambda$}
= $\mathit{\ell}$+ 1 $\forall\ell$, and \textit{$\lambda$} = 0,
that is, $\lambda\in N$. As they are less singular at the origin
than 1/\textit{$r{}^{3}$}, the symbol Pf. can be dropped {[}10{]},
and (13) is written

\begin{equation}
H_{d}\Psi_{n}=E_{n}\Psi_{n}\qquad\qquad\lambda\in N
\end{equation}

In particular for $\ell$ = 0, $\Psi_{n}$ = $(4\pi){}^{-1/2}[u_{n}(r)/r]$,
whether $u_{n}\left(0\right)=0$ if \textit{$\lambda$} = 1, or $u_{n}\left(0\right)\neq0$
if \textit{$\lambda$} = 0, in \textit{both} cases 
\begin{equation}
H_{d}\frac{1}{\sqrt{4\pi}}\frac{u_{n}(r)}{r}=\frac{1}{\sqrt{4\pi}}E_{n}\frac{u_{n}(r)}{r}\qquad\qquad\lambda=0
\end{equation}

where $\mathit{u_{n}\left(r\right)}$ and $\mathit{E_{n}}$ are solutions
of the radial equation (4). $\mathit{H_{d}}$ is thus the \textit{required}
\textit{operator}.

In order to compare the operators $\mathit{H_{d}}$ and\textit{ H},
let us recall that the Hamiltonian of $\textrm{Pf.}\Psi_{n}{\displaystyle {\textstyle =\textrm{Pf}.r^{\lambda-1}\sum_{k\geq0}a_{k}^{n}r^{k}}\mathit{Y_{\ell}^{\mu}\left(\theta,\varphi\right)}}$
is the Hamiltonian in the sense of the distributions given by {[}1{]}

\[
H\textrm{Pf}.\Psi_{n}=\textrm{Pf}.\left\{ \frac{1}{r}\left[-\frac{\hbar^{2}}{2m}\frac{d^{2}}{dr^{2}}+\frac{\ell(\ell+1)\hbar^{2}}{2mr^{2}}+V(r)\right]\mathit{u_{n}}(r)Y_{\ell}^{\mu}\left(\theta,\varphi\right)\right\} -\frac{\hbar^{2}}{2m}Q_{\lambda,\ell}\left(\delta\right)
\]

\[
Q_{\lambda,l}\left(\delta\right)=\sum_{k=0}^{-\lambda}a_{k}\chi_{p(k)}B_{\ell,p(k)}C_{p(k)}\mathit{r{}^{\ell}\mathit{\mathit{Y_{\ell}^{\mu}\left(\theta,\varphi\right)}}}\Delta^{p(k)}\delta\qquad p=-\frac{k+\lambda-\ell}{2}
\]

\begin{equation}
\chi_{p}=\begin{cases}
1\; & if\; p\in N\\
0 & if\; p\notin N
\end{cases}\qquad B_{\ell,p}=\frac{1-2\ell}{4p+1}\qquad C_{p}=-\frac{(4p+1)\pi^{3/2}}{2^{2p-1}p!\,\Gamma(p+3/2)}
\end{equation}

where $\Delta^{p}$ is the iterated Laplacian,$\delta$ the Dirac
mass, where the even, resp. odd, terms of the sum are zero , if \textit{$\lambda-\ell$}
is odd, resp. even, and where $\mathit{Q_{\lambda,\ell}\left(\delta\right)}$
$\neq$ 0 if and only if there is at least one coefficient $a_{k}\neq0$
for which \textit{k} + \textit{$\lambda-\ell$} is an even negative
integer {[}1{]}. As

\begin{equation}
(H_{d}-H)\textrm{Pf.}\Psi_{n}=\left[\textrm{Pf}.,H\right]\Psi_{n}=\frac{\hbar^{2}}{2m}Q_{\lambda,\ell}\left(\delta\right)
\end{equation}

the operators $\mathit{H{}_{d}}$ and \textit{H} differ when they
act on pseudofunctions for which the operators\textit{ }Pf. and \textit{H}
do not commute, or for which $\mathit{Q_{\lambda,\ell}\left(\delta\right)}$$\neq$
0. Considering that energy levels are given in any case by the operator
\textit{H} amounts to identify the operators \textit{H}$_{d}$ and
\textit{H}, and hence to take no account of the noncommutation of
the operators Pf. and \textit{H}. As the operators $\mathit{H{}_{d}}$
and \textit{H} are equivalent in $\textrm{R}^{3}$/\{0\}, and in $\textrm{R}^{3}$
when the Laplacian is the Laplacian in the sense of the functions,
it comes from the confusion between the Laplacians in $\textrm{R}^{3}$/\{0\}
and in $\textrm{R}^{3}$, or/and between the Laplacians in the sense
of the functions and in the sense of the distributions {[}11{]}. By
substituting (10) in (16), we obtain

\begin{equation}
H\textrm{Pf}.\Psi_{n}=E_{n}\textrm{Pf}.\Psi_{n}-\frac{\hbar^{2}}{2m}Q_{\lambda,\ell}\left(\delta\right)
\end{equation}

When the solutions of the radial equation are given by the root \textit{$\lambda$}
= $\mathit{\ell}$+ 1, \textit{p} = \textendash{} (\textit{k }+ 1)/2\textit{
$\notin N$} \textit{$\forall k$, }so that $Q_{\ell+}{}_{1,\ell}$
= 0, $\Psi_{n}$ behaves at the origin like $r{}^{\ell}Y_{\ell}^{\mu}\left(\theta,\varphi\right)$,
it is an eigenfunction of\textit{ H. }When the solutions of the radial
are given by the root $\lambda$ = $-\ell$, as $\mathit{a_{o}\neq}$
0 and \textit{p} =\textit{ }$\ell\in N$ for \textit{k =} 0, $Q{}_{-\ell,\ell}\neq0$,
$\textrm{Pf}.\Psi_{n}$ behaves at the origin like $\textrm{Pf}.r{}^{-(\ell+1)}Y_{\ell}^{\mu}\left(\theta,\varphi\right)$,
it is not an eigenfunction of \textit{H }{[}1{]}. As $Q{}_{0,0}=-\sqrt{4\pi}u_{n}(0)\delta$,
in the case of normalizable solutions (18) is written

\[
H\Psi_{n}=E_{n}\Psi_{n}\qquad\qquad\lambda=\ell+1
\]

\begin{equation}
H\,\frac{1}{\sqrt{4\pi}}\frac{u_{n}(r)}{r}=\frac{1}{\sqrt{4\pi}}E_{n}\frac{u_{n}(r)}{r}+\frac{\hbar^{2}\sqrt{\pi}}{m}u_{n}(0)\delta\qquad\qquad\lambda=0
\end{equation}

Comparison between (14), (15) and (19) shows that \textit{energy levels,
which are given by eigenfunctions of H when $u_{n}$(0) = 0}, \textit{are
given in any case by eigenfunctions of }$\mathit{H_{d}}$.

\section{The Hamiltonian, the Schrödinger operator and the operator $\mathit{H_{d}}$. }

The differential form

\begin{equation}
H=-\frac{\hbar^{2}}{2m}\Delta+V
\end{equation}

is usually called Hamiltonian or Schrödinger operator. Now, a differential
form such as (20) defines an operator only on a given class of functions,
so that the same differential form can define different operators
{[}See e.g. 12{]}. Any function \textgreek{Y}, or pseudofunction Pf.\textgreek{Y},
belongs to the domain $\mathcal{D}$ on which the Hamiltonian, denoted
by \textit{H}, is defined by (20). On the other hand, the operator
involved in the Schrödinger equation, or Schrödinger operator, denoted
by $\mathit{H_{s}}$, can be defined on the set $\mathcal{D}{}_{s}$
of the functions which are any superpositions of eigenfunctions of
\textit{H}. As $\mathcal{D}{}_{s}$ is a subset of $\mathcal{D}$,
and as the two operators are equivalent on $\mathcal{D}{}_{s}$, that
is, $H\Psi\equiv H_{s}\Psi$ if $\Psi\in\mathcal{D}_{s}$, the Schrödinger
operator is a \textit{restriction} of the Hamiltonian. As long as
one confines oneself to solutions of the Schrödinger equation, these
two operators, which have the same eigenfunctions, are equivalent,
so that the distinction is not necessary, and in fact it is not made.
Nevertheless the distinction can be made with respect to the operator
$\mathit{H}_{d}$.

The latter is defined on the same domain $\mathcal{D}$ as \textit{H}.
Moreover (13) and (18) show that the eigenfunctions of \textit{H},
or of $\mathit{H{}_{s}}$, form a subset of the eigenfunctions of
$\mathit{H{}_{d}}$, so that if $\Psi\in\mathcal{D}_{s}$, then $\mathit{H_{d}}\Psi\equiv H_{s}\Psi$.
The operator $\mathit{H_{d}}$ is thus also an extension of the Schrödinger
operator. But the fact that the eigenvalue equation of $\mathit{H{}_{d}}$
is an extension of the eigenvalue equation of \textit{H} to the pseudofunctions
such that $\mathit{Q_{\lambda,\ell}\left(\delta\right)}\neq$ 0 does
not mean that $\mathit{H{}_{d}}$ is an extension of \textit{H}. Eq.(17)
shows indeed that they are different when $\mathit{Q_{\lambda,\ell}\left(\delta\right)}\neq$
0, or when the operators Pf. and \textit{H} do not commute. It follows
that \textit{the operators $\mathit{H}$ and $\mathit{H{}_{d}}$ are
two different extensions of the Schrödinger operator}.

The condition for the operator \textit{$\mathit{H}$} (and then \textit{$\mathit{H}_{s}$})
to be self-adjoint is usually written \textit{u}(0) = 0. However,
according to Merzbacher the condition that the Hamiltonian must be
self-adjoint implies that any two physically admissible eigensolutions
of (2) must satisfy the condition {[}13{]}

\begin{equation}
\textrm{lim }_{r\rightarrow0}\left(u_{1}^{*}\frac{du_{2}}{dr}-u_{2}\frac{du_{1}^{*}}{dr}\right)=0
\end{equation}

This condition is satisfied, either for \textit{u}(0) = 0 and \textit{u}
'(0)$\neq$ 0, in which case $\Psi$ is an eigenfunction of $\mathit{H_{d}}$
and of \textit{H}, or for \textit{u} '(0) = 0 and \textit{u}(0) $\neq$
0, in which case $\Psi$ is an eigenfunction of $\mathit{H_{d}}$,
but not of \textit{H}. It follows that (21) is in fact the condition
for $\mathit{H_{d}}$ to be self-adjoint.

The operator $\mathit{H_{d}}$ can also be defined, by using the linearity
of the operator Pf., from the operator $\textrm{Pf}.\Delta$ introduced
in {[}11{]}. As

\[
\textrm{Pf}.H=\textrm{Pf}.(-\frac{\hbar^{2}}{2m}\Delta+V)
\]
by similarly defining the operator $\Delta_{d}$ as

\begin{equation}
\Delta_{d}\textrm{Pf}.\Psi=\textrm{Pf}.\Delta\Psi
\end{equation}

we have

\[
H_{d}=-\frac{\hbar^{2}}{2m}\Delta_{d}+V
\]

\section{A physically not required condition.}

The boundary condition $\mathit{u_{n}}$(0) = 0 is definitely stated
as a condition required to obtain energy levels because it is the
condition for $\Psi_{n}$ to be an eigenfunction of the Hamiltonian,
but in practice this condition is not respected. It can be imposed
when the solutions are determined with the help of numerical or approximation
methods, but not when energy levels are determined with the help of
analytic solutions. 

In the case of diatomic molecules, in the parabolic like in the Morse
potential, the fact that $\mathit{u_{n}}(0)\neq0$ is regarded as
of no significance owing to the fact that the origin being some distance
outside of the classically allowed region, $\mathit{u_{n}}$(0) is
\textquotedblleft{}very small\textquotedblright{}. As the parabolic
potential (5) is an expansion of the potential in the neighborhood
of the minimum by neglecting the terms of order higher than 2, it
remains close to the physical potential only in the neighborhood of
the minimum. The Morse potential (9) is closer to the physical potential,
and in a region more extended on both sides of the minimum. Comparison
between the measured vibrational levels $\mathcal{E}{}_{n}$ and the
eigenvalues $\mathit{E_{n}}$ of the radial equations involving these
potentials shows that the agreement between these eigenvalues and
the energy levels decreases with \textit{n}, and that the number of
eigenvalues which are close to energy levels is all the greater as
the region in which \textit{V}(\textit{r}) \ensuremath{\approx} $\mathcal{V}$(\textit{r})
is more extended. It means that this agreement essentially depends
on $\mathit{u_{n}}$(\textit{r}) in the region where \textit{V}(\textit{r})
is close to $\mathcal{V}$(\textit{r}), where one therefore expects
that $\mathit{u_{n}}$(\textit{r}) is close to $\mathit{w_{n}}$(\textit{r}),
the region where nearly the whole wave function is concentrated. It
follows that if energy levels are not affected by the fact that $\mathit{u_{n}}$(0)
\ensuremath{\neq} 0, it does not come from the fact that $\mathit{u_{n}}$(0)
is \textquotedblleft{}very small\textquotedblright{}, but from the
fact that the origin is outside of the region where \textit{V}(\textit{r})
$\approx\mathcal{V}(r)$, where $u_{n}\left(r\right)\approx w_{n}\left(r\right)$,
and where the wave function is concentrated, that is, in a region
where $\mathit{u_{n}}$(\textit{r}) has little or none influence on
the fact that $\mathit{E_{n}}$ is close or not to $\mathcal{E}{}_{n}$.

Besides, in the case of the potential (5) $\mathit{u_{n}}$(\textit{r})
always vanishes \textit{inside} of the classically allowed region.
It means in particular that the condition $\mathit{u_{n}}$(0) = 0
can be satisfied only inside of this region.

Consider a physical potential $\mathcal{V}{}_{o}\left(r\right)$,
and a supplementary potential, or a perturbation, $\mathcal{W}\left(r\right)$,
for which one substitutes the spherical oscillator $V_{o}\left(r\right)=\text{\textonehalf}m\omega{}^{2}r^{2}$
and the linear potential \textit{W}(\textit{r}) = \textendash{} \textit{Cr},
so that the approximation of the physical potential $\mathcal{V}\left(r\right)$
= $\mathcal{V}{}_{o}\left(r\right)$ + $\mathcal{W}\left(r\right)$
is the potential \textit{V}(\textit{r}) = $V_{o}\left(r\right)$ +
\textit{W}(\textit{r}) of the form (5), that is,

\[
V\left(r\right)=\text{\textonehalf}m\omega{}^{2}r^{2}-Cr=\text{\textonehalf}m\omega{}^{2}\left(r-r_{m}\right)^{2}+V_{m}
\]

\begin{equation}
r_{m}=C/m\omega^{2}\qquad\qquad V_{m}=C^{2}/2m\omega^{2}
\end{equation}

As \textit{C} and then $\mathit{r{}_{m}}$ have any values, the origin
is or not outside of the classically permitted region. When $\beta\mathit{r_{m}}$
coincides with a zero of a Hermite polynomial of order \textit{N},
that is, $\mathit{H_{N}\left(\beta r_{m}\right)}$ = 0, then $u_{N}\left(0\right)$
= 0, but $u_{n}\left(0\right)$ $\neq$ 0 for \textit{n} < \textit{N}.
Insofar as the agreement between the eigenvalues of the radial equation
and the energy levels decreases with \textit{n}, the eigenvalues $\mathit{E_{n}}$
with \textit{n} < \textit{N} that belong to solutions $u_{n}\left(r\right)$
which do not satisfy the condition $u_{n}\left(0\right)$ = 0 must
be closer to an energy level than $\mathit{E_{N}}$ that belongs to
a solution $u_{N}\left(r\right)$ which satisfies this condition.
The latter then cannot be considered as the condition required to
obtain accurate, or better estimates of energy levels.

In the absence of supplementary potential, or of perturbation, the
approximation of the physical potential is the spherical oscillator,
$\mathcal{V}$(\textit{r}) \ensuremath{\approx} $V$(\textit{r}),
with
\[
V\left(r\right)=\text{\textonehalf}m\omega{}^{2}r^{2}
\]

\[
u_{n}\left(r\right)=N_{n}exp(-\text{\textonehalf}\beta^{2}r^{2})H_{n}(\beta r)\qquad E_{n}=(n+\text{\textonehalf)}\hbar\omega
\]

\[
u_{n}\left(0\right)=N_{n}H_{n}(0)
\]

As Hermite polynomials satisfy the condition $H_{2p}\left(0\right)\neq0$,
$H_{2p+1}\left(0\right)=0$, half the square-integrable solutions
of the radial equation satisfy the condition $\mathit{u_{n}}$(0)
= 0, and hence half the energy levels are given by eigenvalues of
\textit{H}. If the supplementary potential, or the perturbation, is
a 1D potential $\mathcal{W}$(\textit{x}) for which one substitutes
the linear potential \textit{W}(\textit{x}) = \textendash{} \textit{Cx},
the potential \textit{V}(\textit{r}) = \textit{$V{}_{o}(r)$} +\textit{
W}(\textit{x}) used as an approximation of the physical potential
$\mathcal{V}$(\textit{r}) = $\mathcal{V}{}_{o}$(\textit{r}) + $\mathcal{W}$(\textit{x})
is still a spherical oscillator, that is,

\[
V\left(\rho\right)=\text{\textonehalf}m\omega{}^{2}r^{2}-Cx=\text{\textonehalf}m\omega{}^{2}\rho^{2}-V_{m}
\]

\[
\rho=\left[(x-C/m\omega^{2})^{2}+y^{2}+z^{2}\right]\qquad V_{m}=C^{2}/2m\omega^{2}
\]

Half the energy levels are still given by eigenvalues of \textit{H}.
But when the supplementary potential, or the perturbation, is the
3D potential $\mathcal{W}$(\textit{r}) for which one substitutes
the linear potential \textit{W}(\textit{r}) = \textendash{} \textit{Cr},
the potential used as an approximation of the physical potential is
the potential (23), so that the solutions of the radial equation,
which are given by (7), do not satisfy the condition $\mathit{u_{n}}$(0)
= 0, and then that the energy levels are not given by eigenvalues
of \textit{H}, as it is the case for diatomic molecules. It shows
that whether or not energy levels are given by eigenvalues of \textit{H}
only depends on the potential used as an approximation of the physical
potential \textendash{} and hence on the latter.

Perturbation methods, in particular variational methods, apply to
the eigenfunctions of \textit{H} on which the theory is based. By
substituting $\mathit{H_{d}}$ for \textit{H} these methods are extended
to the eigenfunctions of $\mathit{H_{d}}$ which behave at the origin
like 1/\textit{r}. It is in agreement with the fact that energy levels
are obtained with the help of such functions. Similarly, situations
which are described with the help of functions which behave at the
origin like 1/\textit{r}, require, to hold in $\textrm{R}{}^{3}$,
and not only in $\textrm{R}{}^{3}$/\{0\}), that one substitutes $\mathit{H_{d}}$
for \textit{H}. This is the case of ingoing, outgoing and standing
spherical waves which are described with the help of Green functions.
As the latter behave at the origin like 1/\textit{r}, they are eigenfunctions
of \textit{H} in $\textrm{R}{}^{3}$/\{0\}), but not in $\textrm{R}{}^{3}$,
so that this description does not hold at the origin, which is physically
unsatisfactory. As Green functions are eigenfunctions of $\mathit{H_{d}}$
in $\textrm{R}{}^{3}$, by substituting this operator, the description
of spherical waves in terms of Green functions holds in the whole
space.

Moreover, wave functions divergent at the origin, but normalizable,
are encountered in the relativistic theory of the hydrogen atom. As
the relativistic generalization of the Schrödinger equation is the
Klein-Gordon equation which governs the wave function in the absence
of electromagnetic field for a relativistic particle with spin zero,
if the Laplacian must be taken in the sense of the distributions in
this equation, one will be similarly led to substitute for $\Delta$
the operator $\Delta_{d}$ defined in (22).

\section{Conclusion. }

One usually considers, by putting forward different arguments, that
the resolution of the Schrödinger equation must be supplemented with
the boundary condition $\mathit{u_{n}}$(0) = 0, but when this condition
is not satisfied, one admits, by putting forward other arguments,
that this is of no significance. In fact the resolution of the Schrödinger
equation need not be supplemented with any boundary condition {[}1{]},
but the fact that the condition $\mathit{u_{n}}$(0) = 0 is not satisfied
is significant, since it means, whether or not $\mathit{u_{n}}$(0)
is very small, that energy levels are not given by solutions of the
Schrödinger equation. As energy levels are obtained by substituting
theoretical and then approximate potentials, it should be taken for
granted that in approximate solutions the boundary condition $\mathit{u_{n}}$(0)
= 0, and hence the operator \textit{H}, or $\mathit{H_{s}}$, are
not required to obtain energy levels, and hence that the operator
$\mathit{H{}_{d}}$, less stringent, is sufficient. The point is that
the only potentials that are exactly known, and then used to obtain
energy levels, are theoretical and then approximate potentials.

\end{document}